\documentclass[aps,pre, twocolumn,
              groupedaddress,showpacs,showkeys]{revtex4-1}
\usepackage[dvips]{graphicx}
\usepackage{cmap}
\usepackage{epstopdf}
\begin{document}

% Use the \preprint command to place your local institutional report
% number in the upper righthand corner of the title page in preprint mode.
% Multiple \preprint commands are allowed.
% Use the 'preprintnumbers' class option to override journal defaults
% to display numbers if necessary
%\preprint{}

%Title of paper
\title{Power-law multi-wave model for COVID-19 propagation \\ in countries with nonuniform population density}

% repeat the \author .. \affiliation  etc. as needed
% \email, \thanks, \homepage, \altaffiliation all apply to the current
% author. Explanatory text should go in the []'s, actual e-mail
% address or url should go in the {}'s for \email and \homepage.
% Please use the appropriate macro foreach each type of information

% \affiliation command applies to all authors since the last
% \affiliation command. The \affiliation command should follow the
% other information
% \affiliation can be followed by \email, \homepage, \thanks as well.
\author{P.S. Grinchuk, S.P. Fisenko}
\email[P.S. Grinchuk $\:$]{gps@hmti.ac.by}
\homepage[]{www.itmo.by} 
%\thanks{}
%\altaffiliation{}
\affiliation{A.V.Luikov Heat and Mass Transfer Institute, National Academy of Sciences of
Belarus,\\ 15 P. Brovka Str., Minsk 220072, Belarus.}

%Collaboration name if desired (requires use of superscriptaddress
%option in \documentclass). \noaffiliation is required (may also be
%used with the \author command).
%\collaboration can be followed by \email, \homepage, \thanks as well.
%\collaboration{}
%\noaffiliation

\date{\today}

\begin{abstract}
The phenomenological mathematical model of COVID-19 spreading is proposed for large countries with geographical differentiation of population density. According to the model COVID-19 spreading takes the form of several spatio-temporal waves developing almost independently and simultaneously in areas with different population density. The intensity of each wave is described by a power-law dependence. 
The parameters of dependence are determined by real statistical data at the initial stage of the disease spread. The results of the model simulation were verified using statistical data for the Republic of Belarus. Based on the developed model, a forecast calculation was made at the end of May, 2020. The accuracy of forecasting the total number of cases for a period of 3 months in the proposed approach was about 3\%.
\end{abstract}

% insert suggested PACS numbers in braces on next line
% insert suggested keywords - APS authors don't need to do this

%\maketitle must follow title, authors, abstract, \pacs, and \keywords
\maketitle

%%%%%%%%%%%%%%%%%%%%%%%%%%%%%%%%%%%%%%%%%%%%%%%%%%%%%%%%%%%%%%%%%%%%%%%%%
\section*{Introduction}
Predicting the spread of the coronavirus COVID-19 is one of the most challenging problems for the relevant scientific communities of medics, mathematicians and physicists \cite{Ref1,Ref1a}. 
An accurate forecast allows to correctly allocating available resources to effectively withstand the epidemic \cite{Ref1b,Ref_gps-dts}. 
It is impossible to predict the spreading of COVID-19 with absolute accuracy. 
However, forecasting can be improved by adjusting the mathematical models used and comparing the forecasting data with the actual situation.

A characteristic feature of the development of the epidemic situation in some of countries was 
a long plateau in the number of new cases of the COVID-19 disease or a slow decrease in daily cases in April-May 2020 \cite{Ref2}. 
This was observed in the United States (April and May), United Kingdom and Canada (from the beginning of April to the middle of May), Russia (in May), Belarus (from the last decade of April to the last decade of May), Sweden (April and May), Indonesia (April and May), Poland (April and May), Ukraine (April and May). 
To note that this situation with long plateau is characteristic for the total epidemic indicators worldwide, when the total number of sick people is at the level of 80-100 thousand new cases per day starting from the first decade of April to the last decade of May.

For fast diseases, with an incubation period of several days \cite{Ref3}, the presence of a long plateau is not typical situation. 
More typical situation is the epidemic development according to the scenario ''growth - peak phase - decline'', when the duration of the peak phase is several times shorter than the growth phase \cite{Ref4}. 
For the cases under discussion, the plateau duration is comparable and even exceeds the duration of the rapid growth phase (approx. 30 days and more). This feature raises questions about our understanding of this phenomenon.

To the middle of spring 2020, there were reports about the expectation of the second wave of the COVID-19 spread \cite{Ref4}. 
The development of the second epidemic wave in many countries in the fall of 2020 is no longer in doubt.
%–азвитие второй волны эпидемии во многихх странах мира осенью 2020 года уже не вызывает сомнений. 
%Something similar to the second wave in May and beginning of June 2020 developed in Iran \cite{Ref2}. 
Note that this refers to the second wave of the disease in the same population in some time. 
But we consider the phenomenon of the simultaneous propagation of several waves in different subpopulations of the same population. 
This phenomenon can explain the current situations with long plateau. Below is presented the mathematical model that describes this phenomenon.

% % % % % % % % % % % % % % % % % % % % % % % % % % % % % % % % % % % % % % %

\section{Mathematical model}

As a result of the analysis of the possible reasons of the long plateau phenomenon, we came to the conclusion that it can be associated with regional characteristics of the nonuniform distribution of the population density and the formation of the additional waves of the virus decease spreading in corresponding subpopulations. 
To note that in this work we do not consider well known effect of the presence of a large number of asymptomatic cases of a mild course of the COVID-19 \cite{Ref4}. 
In fact, such cases are not amenable to counting and for analysis in real epidemic situation. We will only use information about reported cases, which are usually associated with more severe manifestations of the disease. 
Using the open statistics for the Republic of Belarus \cite{Ref2}, we want to show that a long plateau in this case can be described by the simultaneous propagation of three independent waves of a COVID-19 disease in the country. 

The essence of this phenomenon, in our opinion, is as follows. At the beginning of the epidemic, all countries closed their borders. 
Therefore, each of the countries, including the Republic of Belarus, can be considered as a closed and isolated human community. 
However, within many countries the population density is nonuniform one. 
It can be considered as quite reliable fact that a population density is one of the key factors influencing the spread of the respiratory virus \cite{Ref7,Ref8}. 
In large cities with a high density, the spread of the virus is most intense. 

From the point of view of population density, the Republic of Belarus can be divided into 3 agglomerations: 
1) the Minsk city with the population density of 5800 ppl/km$^2$;
2) regional centers, where the population density is from 2400 ppl/km$^2$ (Brest), 2600 ppl/km$^2$ (Grodno) to 3800 ppl/km$^2$ (Gomel), and large cities of regional subordination (Polotsk, Orsha, Bobruisk, Pinsk, etc.), with a population density of 2100 ppl/km$^2$ (Polotsk) to 2950 ppl/km$^2$ (Orsha); 
3) medium and small district centers and rural settlements with a population density of tens to hundreds of people per square kilometer. 

It is worthy to note that the average population density for the Republic of Belarus is 46 ppl/km$^2$. 
The spread of the virus in each of these agglomerations should occur at its own speed, the onset of the disease in each of the agglomerations can also be shifted in time \cite{Ref8}. 
We assume that the spread of the disease begins in the largest city with the highest population density and then, with a certain delay, the disease comes in an agglomeration with a lower density. 
In the first approximation, it can be considered that the disease in each of the agglomerations proceeds independently and mainly propagates within a subpopulation. 
These considerations underlie the analysis below.

% % % % % % % % % % % % % % % % % % % % % % % % % % % % % % % % % % % % % % % % % % % % % 
%11111111111111111111111111111111111111
% For one-column wide figures use
\begin{figure}[pb]
% Use the relevant command to insert your figure file.
% For example, with the graphicx package use
\includegraphics[width=\linewidth]{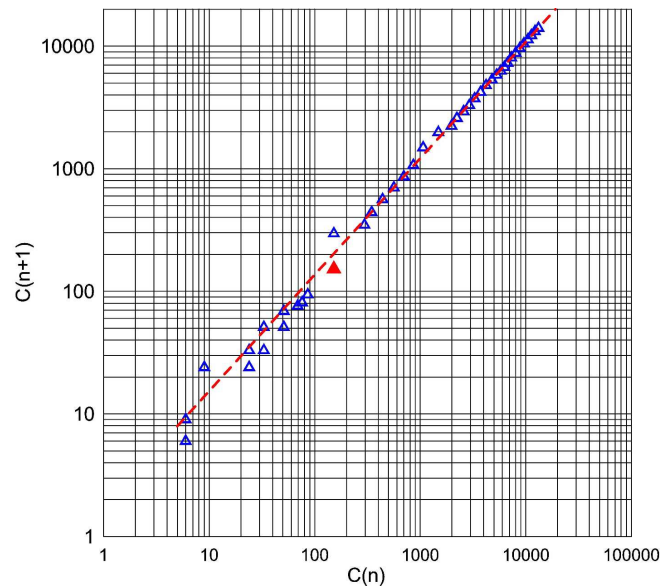}
\vspace*{8pt}
% figure caption is below the figure
\caption{The dependence of the number of $C(n+1)$ patients involved in the disease on day $n+1$ on the number of $C(n)$ people involved on the previous day according to the statistics of the incidence of coronavirus in the Republic of Belarus from March 6 to April 30, 2020. The approximation performed for the period from March 30 (red triangle) till April 30}
\label{fig:1}       % Give a unique label
\end{figure}
% % % % % % % % % % % % % % % % % % % % % % % % % % % % % % % % % % % % % % % % % % % % % % 

Previously, we used the described mathematical approach\cite{Ref9} to analyze and forecast the situation with the dynamics of the coronavirus spreading in the Republic of Belarus \cite{Ref10,Ref11,Ref12}. 
The essence of the analysis is to graph the dependence of new cases of the disease on cases in the previous day in double logarithmic coordinates (Figure \ref{fig:1}), 
\cite{Ref10}. The authors of Ref.~\cite{Ref9} found a general pattern for a number of countries, namely, the power-law dependence of the number of new cases on a given day n+1 on the number of cases on the previous one:

\begin{equation}
C(n+1)=\alpha C(n)^{\beta}.
\label{eq1}
\end{equation}

Here $n$ is the days counted from the beginning of the spread of the epidemic disease. 
%ќтметим, что причина хорошего описани€ эпидемической ситуации соотношением (1) до конца не пон€тна. Ёто требует дополнительного исследовани€. ѕоскольку это соотношение лежит в основе нашего рассмотрени€, то предложенную модель мы назвали феноменологической .
The reason for the good description of the epidemic situation by relation (\ref{eq1}) is not fully understood. 
This requires additional research. 
Since this relationship underlies our mathematical consideration, we called the proposed model phenomenological.
The exponent $\beta$ in expression (\ref{eq1}) and the proportionality coefficient $\alpha$ should remain constant over time if the intensity of social contacts and the scenarios of social behavior in the population did not change. 
The presence of such dependence makes it possible to determine its parameters at the initial stage of the epidemic and to forecast a change in the disease rate over quite long time periods.

It can be shown\cite{Ref8} that the equation (\ref{eq1}) leads to a geometric progression,
which in turn gives an analytical expression for the number of diseased people $C(n)$ over time, i.e. from the number of days $n$ from the start of the count:

%It can be shown \cite{Ref8} that equation (\ref{eq1}) leads to a certain analytical %expression for the number of diseased people $C(n)$ versus time, i.e. from the number of %days $n$ from the start of the count:

\begin{equation}
C(n)=\alpha^{\frac{1-\beta^n}{1-\beta}} C(0)^{\beta^n}.
\label{eq2}
\end{equation}

To build a forecast curve, one also needs the initial value of the number of involved people $C(0)$. 
It is usually accepted as the starting day to use the date when the total number of cases in the country has reached a certain level. 
Since the disease statistics always show significant fluctuations at the initial stage, dependence (\ref{eq1}) is applicable when the number of cases is about 100 people or more. Initially the parameter $C(0)$ in our work was determined from the condition of the minimum dispersion in approximation of the actual data in the period from March 30 to April 24:

\begin{equation}
C(0) \approx 100.
\label{eq3}
\end{equation}

As noted above, the day when the total number of cases exceeded 100 is selected as the starting point for our simulation. For the Republic of Belarus, this is March 30$^{th}$, 2020.

The considered mathematical approach is in agreement with the logistic equation of the spreading of virus diseases\cite{Ref13,Ref14,Ref15}. 
The equation of type (\ref{eq1}) for the dynamics of the viral disease spread is closely related to the classical approach based on the Verhlust logistic equation. 
In similar notation, this equation can be written as\cite{Ref16}

\begin{equation}
\frac{dC(t)}{dt}= k b C(t) \left\lbrace  1- \frac{C(t)}{N} \right\rbrace.
\label{eq4}
\end{equation}

Here $k$ is the probability of transmission of the infection from the patient to a healthy person, averaged over the entire population, $b$ is the average number of contacts per day for the average person, $N$ is the population size (population of the country), $t$ is temporal variable.
Note that the well-known viral disease reproductive number $R_0$ \cite{Ref17} is written in these notation as $R_0 = k b$. 
The solution of equation (\ref{eq4}) with the initial condition  
$C |_{t=0}= C(0)$  is well known and has the form

\begin{equation}
C(t) = \frac{N C(0)\exp(R_0 t)}{N+C(0) [\exp(R_0 t) - 1)]}.
\label{eq5}
\end{equation}

Then, for short times, $t\ll t_{*} = 1/R_0$, the number of cases should grow exponentially

\begin{equation}
C(t) \approx C(0)\exp(R_0 t).
\label{eq6}
\end{equation}

Comparing equations (\ref{eq2}) and (\ref{eq6}) and taking into account that the small parameter is $1-\beta \approx 0.05$, we can establish an approximate relation between the parameters

\begin{equation}
R_0 = k b \approx \ln \alpha + (\beta - 1) \ln C(0), \quad t,n \ll t_*.
\label{eq7}
\end{equation}

%\begin{figure}[h!] 
%\begin{center} 
%\includegraphics[width=\linewidth]{transport} 
%\caption{...} 
%\end{center} 
%\end{figure}

%22222222222222222222222222222222222222
\begin{figure}[h!]
% Use the relevant command to insert your figure file.
% For example, with the graphicx package use
\includegraphics[width=\linewidth]{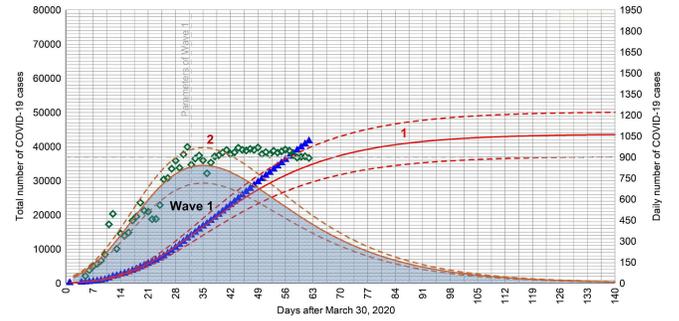}
\vspace*{8pt}
% figure caption is below the figure
\caption{Results of forecast calculating the dynamics of the COVID-19 spreading in the Republic of Belarus in the approximation of a single propagation wave. %\cite{Ref10}. 
The wave parameters were determined as of April 30, 2020 (equation (3.1)). 
Actual data are shown until May 30. Curve 1 (left ordinate) is forecast calculation for the total number of disease cases; blue characters is actual data; curve 2 (right axis of ordinates) is forecast calculation for the number of daily cases of disease, green symbols is actual data. Dashed lines define the area of forecast calculations error}
\label{fig:2}       % Give a unique label
\end{figure}
%
%3333333333333333333333333333333333333
\begin{figure}
% Use the relevant command to insert your figure file.
% For example, with the graphicx package use
\includegraphics[width=\linewidth]{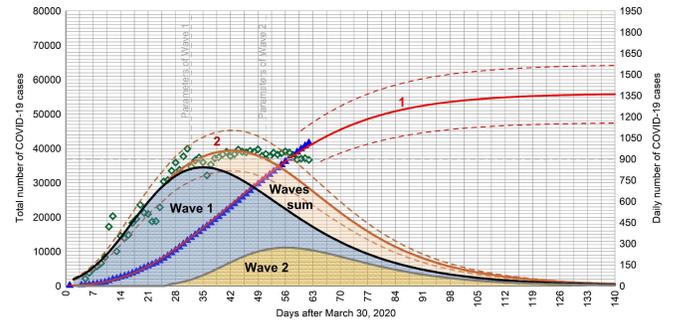}
\vspace*{8pt}
% figure caption is below the figure
\caption{Results of forecast calculating the dynamics of the COVID-19 spreading in the Republic of Belarus in the approximation of two waves %\cite{Ref11}. 
The second wave parameters were determined as of May 19, 2020 (equation (3.2)). 
Actual data are shown until May 30. Curve 1 (left ordinate) is forecast calculation for the total number of disease cases; blue characters is actual data; 
curve 2 (right axis of ordinates) is forecast calculation for the number of daily cases of disease, green symbols is actual data. Dashed lines define the area of forecast calculations error}
\label{fig:3}       % Give a unique label
\end{figure}
% % % % % % % % % % % % % % % % % % % % % %% % % % % % % % % % % % % % % % % % % % % %

According to the Eq. (\ref{eq7}) an increasing in values of the parameters $\alpha$ and $\beta$ indicate a higher reproduction rate in the corresponding population.

Additional data accumulated to mid-May showed a certain deviation from the forecast curve
based on the single wave model \cite{Ref10,Ref11} (Eqs. (\ref{eq1}), (\ref{eq2}); Figure \ref{fig:2}).
Initially, we assumed that the observed morbidity situation can be described more precisely if its propagation is allowed in the form of two waves. 
These waves independently but simultaneously arise, develop and propagate in the two largest population agglomerations described above. 
The addition of the second wave to the model well improved the compliance between the calculation results and the actual data in the period from 40 to 50 days from the start of the count. But then again, a certain deviation occurred in the daily cases of the COVID-19 disease (Figure \ref{fig:3}). 
This suggested that, nevertheless, we are dealing with three simultaneous and independent waves in the three subpopulations described above\cite{Ref12}.

The described nature of the development of three waves of COVID-19 can be confirmed by their initial time shift relative to each other. To obtain the numerical parameters of the model, the following algorithm was proposed. 

The characteristics of the first, most intense, wave can be described by parameters obtained from data at the initial stage of the development of the epidemic (March 6 - April 30). These were done in the preprint\cite{Ref10}. 
New data and their deviation from the first forecast curve for one propagation wave can be used to identify the parameters of the second wave. 
For this, data from 33 to 50 days were used\cite{Ref11} (Figure \ref{fig:3}). 
Finally, additional data from 51 to 57 days were used to identify the parameters of the third wave\cite{Ref12}.  

Thus, we put forward the hypothesis that the epidemic situation with COVID-19 in the Republic of Belarus can be described by three independent waves:

\begin{equation}
C(n) = C_1(n)+C_2(n)+C_3(n).
\label{eq8}
\end{equation}

For the first wave, according to the above assumptions, the following relations are true:

\begin{equation}
C_1(n+1)=\alpha_1 C_1(n)^{\beta_1}.
\label{eq9}
\end{equation}

\begin{equation}
C_1(n)=\alpha_1^{\frac{1-\beta_1^n}{1-\beta_1}} C_1(0)^{\beta_1^n}.
\label{eq10}
\end{equation}

The second wave is described by the following relationships:

\begin{equation}
C_2(n_2+1)=\alpha_2 C_2(n_2)^{\beta_2},
\label{eq11}
\end{equation}

\begin{equation}
C_2(n_2)=\alpha_2^{\frac{1-\beta_2^{n_2}}{1-\beta_2}} C_2(0)^{\beta_2^{n_2}},
\label{eq12}
\end{equation}

\begin{equation}
n_2 = n - \Delta n_2.
\label{eq13}
\end{equation}

Finally, the third wave is described by similar relations:

\begin{equation}
C_3(n_3+1)=\alpha_3 C_3(n_3)^{\beta_3},
\label{eq14}
\end{equation}

\begin{equation}
C_3(n_3)=\alpha_3^{\frac{1-\beta_3^{n_3}}{1-\beta_3}} C_3(0)^{\beta_3^{n_3}},
\label{eq15}
\end{equation}

\begin{equation}
n_3 = n - \Delta n_3.
\label{eq16}
\end{equation}

Here $\Delta n_2$, $\Delta n_3$ is the delay time of the second and third waves relative to the first (in days). 

% % % % % % % % % % % % % % % % % % % % % % % % % % % % % % % % % % % % % % % % % % % %

\section{Results and discussion}

Thus, according to the available data and proposed model (\ref{eq8}) - (\ref{eq16}), it is necessary to determine 11 parameters for the first, second and third waves: $\alpha_1$,
$\beta_1$, $C_1(0)$, $\alpha_2$, $\beta_2$, $C_2(0)$, $\Delta n_2$ and $\alpha_3$, $\beta_3$, $C_3(0)$, $\Delta n_3$.  
As the objective function, the standard deviation of the calculated data and actual ones on the total number of cases for the entire period was considered.

The search for optimal parameters was carried out using the gradient method of the steepest descent \cite{Ref18}. The results of the search are: 

\begin{equation}
\alpha_1 = 1.72(4), \beta_1 =0.94(9), C_1(0)=100,
\label{eq17}
\end{equation}

\begin{equation}
\alpha_2 = 1.72(7), \beta_2 =0.94(2), C_2(0)=24, \Delta n_2= 25,
\label{eq18}
\end{equation}

\begin{equation}
\alpha_3 = 1.72(7), \beta_3 =0.94(2), C_3(0)=9, \Delta n_3= 39.
\label{eq19}
\end{equation}

%4444444444444444444444444444444444444
\begin{figure}
% Use the relevant command to insert your figure file.
% For example, with the graphicx package use
\includegraphics[width=\linewidth]{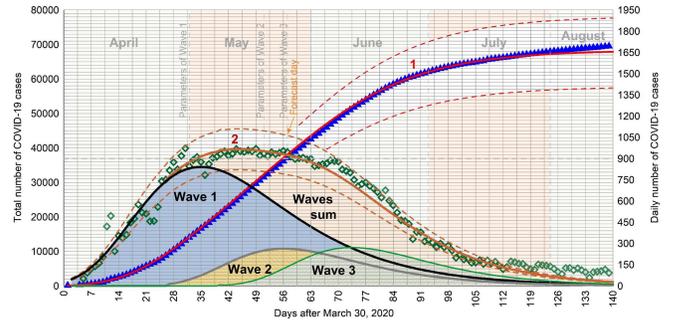}
\vspace*{8pt}
% figure caption is below the figure
\caption{Results of forecast calculating the dynamics of the COVID-19 spreading in the Republic of Belarus in the approximation of three propagation waves. % \cite{Ref12}. 
The third wave parameters were determined as of May 25, 2020 (equation (3.3)). 
Actual data are shown until August 15, 2020. Curve 1 (left ordinate) is forecast calculation for the total number of disease cases; blue characters is actual data; 
curve 2 (right axis of ordinates) is forecast calculation for the number of daily cases of disease, green symbols is actual data. Dashed lines define the area of error of forecast calculations. The color shows the contribution of each of the three waves discussed in the text to the overall dynamics of the disease spread}
\label{fig:4}       % Give a unique label
\end{figure}
% % % % % % % % % % % % % % % % % % % % % % % % % % % % % % % % % % % % % %

It is important to note that the second and third waves are described by identical parameters $\alpha$ and $\beta$, which are similar in magnitude to the parameters of the first wave. 
The difference is only in the third decimal place. 
All parameters coincide within the error of the data processing method used. The second and third waves differ, as we expected, with a lower initial value of the incident cases $C_2 (0)=24$, $C_3 (0)=9$,  and are characterized by a shift of approximately 25 days and 39 days relative to the first wave. 
In a first approximation, it can be assumed that the internal patterns of the distribution of the three epidemic waves are approximately the same (parameters $\alpha$ and $\beta$). 
Waves differ only in their intensity (amplitude) and onset time.

The results of the predictive calculation in the three-wave approximation\cite{Ref12} by the relations (\ref{eq8}) - (\ref{eq19}) are presented in Figure 4. 
The error in our calculations is preliminary estimated at 10-15 \% at time intervals of about 1 months.
The data shown in Figures 2, 3 indicate the correctness of such estimates.  In fact the forecast accuracy is about 3\% for the total number of cases and no more than 15\% for daily new cases within 3 months from the date of the forecast. 
The performed forecast calculations are valid as long as the conditions for the development of the morbidity process correspond to the conditions under which the initial statistics for the forecast were collected.

For a better understanding of the accuracy of various approximations, we presented the results of consideration in the approximation of one, two, and three waves (Figures \ref{fig:2}, \ref{fig:3}, \ref{fig:4}, respectively). 
For clarity, the contribution of each wave is highlighted in color. As can be seen, the most accurate is the approximation of the three-wave model (Figure \ref{fig:4}). 
From the results obtained above, we can conclude that the puzzle of the long plateau and general behavior in COVID-19 dynamics is explained by our phenomenological model. 
But the construction of such a model requires knowledge of the regional characteristics of the country and details of its population density.
% % % % % % % % % % % % % % % % % % % % % % % % % % % % % % % % % % % % % % % % % % % % %

\section{Conclusions}
\label{sec:3}
The phenomenological model is proposed that describes the dynamics of the spread of COVID-19 in Belarus, a country with population about 10 million people. 
We used the approximation of the propagation of three independent waves in subpopulations with different population densities. The dynamics of each of the waves is described by a universal power-law dependence, first discovered in Ref.~\cite{Ref9}. 
The forecast calculations performed using the proposed model on the base of statistical data for the Republic of Belarus showed a high forecast accuracy. 
The model is based on limited statistical information on daily morbidity from March 6 to May 25 (total 81 points, 57 points were used for analysis). 
Nevertheless, the forecast accuracy is about 3\% for the total number of cases and no more than 15\% for daily new cases within 3 months from the date of the forecast. 

We hope that the proposed approaches will be useful for analysis of the pandemic situation in other countries. It is possible that for large countries where the distance between large cities exceeds 500-600 km (the average distance available for a one-day trip by car), 
the propagation dynamics of COVID-19 also will be described by the superposition of a several simultaneous spatio-temporal waves. 
The proposed approach can be applied also for analysis of a spread of seasonal respiratory viral infections, such as influenza.

\section*{Acknowledgment}
This work was supported by the Belarussian Republican Foundation for Fundamental Research.

% % % % % % % % % % % % % % % % % % % % % % % % % % % % % % % % % % % % % % % % % % % % %

% Create the reference section using BibTeX:
%\bibliography{Evaporation_Ref}

\end{document}